\documentclass[preprint,pre,aps]{revtex4-1}

\usepackage{graphicx}

\begin{document}

\title{An alternative derivation of orbital-free density functional theory}
\author{Russell B. Thompson}
\email{thompson@uwaterloo.ca}
\affiliation{Department of Physics \& Astronomy and Waterloo Institute for Nanotechnology, University of Waterloo, 200 University Avenue West, Waterloo, Ontario, Canada N2L 3G1}
\date{May 28, 2019}

\begin{abstract}
Polymer self-consistent field theory techniques are used to derive quantum density functional theory without the use of the theorems of density functional theory. Instead, a free energy is obtained from a partition function that is constructed directly from a Hamiltonian, so that the results are, in principle, valid at finite temperatures. The main governing equations are found to be a set of modified diffusion equations, and the set of self-consistent equations are essentially identical to those of a ring polymer system. The equations are shown to be equivalent to Kohn-Sham density functional theory, and to reduce to classical density functional theory, each under appropriate conditions. The obtained non-interacting kinetic energy functional is, in principle, exact, but suffers from the usual orbital-free approximation of the Pauli exclusion principle in additional to the exchange-correlation approximation. The equations are solved using the spectral method of polymer self-consistent field theory, which allows the set of modified diffusion equations to be evaluated for the same computational cost as solving a single diffusion equation. A simple exchange-correlation functional is chosen, together with a shell-structure-based Pauli potential, in order to compare the ensemble average electron densities of several isolated atom systems to known literature results. The agreement is excellent, justifying the alternative formalism and numerical method. Some speculation is provided on considering the time-like parameter in the diffusion equations, which is related to temperature, as having dimensional significance, and thus picturing point-like quantum particles instead as non-local, polymer-like, threads in a higher dimensional thermal-space. A consideration of the double-slit experiment from this point of view is speculated to provide results equivalent to the Copenhagen interpretation. Thus the present formalism may be considered as a type of ``pilot-wave'', realist, perspective on density functional theory.
\end{abstract}

\maketitle

\section{Introduction}

Density functional theory (DFT) is a huge field that forms a central theoretical pillar of modern materials physics and quantum chemistry \cite{Hohenberg1964, vonBarth2004, Becke2014, Jones2015}. In DFT, the ground state energy of a system is expressed in terms of a functional of the electron density. The most widely used variant of DFT is Kohn-Sham theory (KS) \cite{Kohn1965}, where the energy functional is actually dependent on orbitals which are in turn determined from the density. In this approach, essentially all approximations are contained in the so-called exchange-correlation functional, and so most research is rightly devoted to exploring this functional. A disadvantage of KS-DFT is that a set of eigenvalue equations must be solved, one equation for every explicit electron in the system. 

An alternative is the computationally more efficient ``orbital-free'' (OF) DFT in which only one eigenvalue equation, or the equivalent, needs to be solved \cite{Wang2000, March2010, Karasiev2014, Finzel2017, Carter2018}. This approach, which dates from the ideas of Thomas, Fermi and Dirac \cite{Thomas1927, Fermi1927, Fermi1928, Dirac1930}, is philosophically more consistent with the original spirit of DFT, since it works with functionals of the density directly. A disadvantage of OF-DFT is that, in addition to approximating the exchange-correlation functional, the non-interacting kinetic energy functional, which in KS-DFT is exact, must now be approximated. This is a significant drawback because while the exchange-correlation term is typically the smallest contribution to the energy, the kinetic energy is generally large. 

The purpose of the present paper is neither to improve the exchange-correlation functional nor the kinetic energy. Rather it is to derive a different set of DFT equations from first principles without the use of the theorems of DFT. The resulting equations are valid for finite temperatures, although here they will only be solved for temperatures approaching the ground state. Polymer self-consistent field theory (SCFT) techniques will be used to go from a Hamiltonian to a partition function and from the partition function to a free energy functional. The method is first principles and the resulting equations are effectively identical to ring-architecture polymers \cite{Kim2012}. Standard methods of numerical solution from polymer SCFT can therefore be applied, and the resulting set of modified diffusion equations can, in a spectral representation, be solved for the same computational cost as a single linear polymer. A simple exchange-correlation functional is used, together with a shell-structure-based Pauli potential, in order to benchmark this alternative theoretical and numerical approach against an existing work for (ensembles of) isolated atoms \cite{Finzel2015}. This approach is effectively an OF-DFT related method, but it is shown to become equivalent to KS-DFT under appropriate circumstances. It is also shown to reduce to classical DFT in the appropriate limit. 

A possible quantum interpretation of the SCFT equations is discussed since the same results can be derived from classical statistical mechanics, using SCFT for ring polymers \cite{Kim2012}. It is speculated that by taking an ensemble interpretation of quantum mechanics, and replacing the wave function postulate by one in which quantum particles are considered as thermal Gaussian threads in a four dimensional thermal-space, as suggested by the temperature-related time-like independent-variable of the SCFT governing diffusion equation, that a pilot-wave related theory for quantum mechanics results. Since the crucial modified diffusion equation can be experimentally verified, this offers a potential quantum interpretation that would not be in contradiction with any predictions of standard Copenhagen quantum mechanics.

\section{Theory}

The following derivation of the DFT functional and SCFT equations will be done in the canonical ensemble. Normally, DFT is approached using the theorems of DFT \cite{Hohenberg1964} in the grand canonical ensemble with a constraint to force the chemical potential to give the correct number of particles \cite{Finzel2015}. Since the number of particles is often known, the canonical ensemble can be a more convenient choice and, when working from a first principles partition function without the theorems of DFT, it is easier to use.

Consider $N$ quantum mechanical particles in a canonical ensemble of volume $V$ and temperature $T$. The quantum Hamiltonian can be written in terms of the kinetic energy $\mathcal{K}$ and potential energy $\mathcal{U}$ as
\begin{eqnarray}
\mathcal{H} &=& \mathcal{K} + \mathcal{U} \nonumber \\
&=& -\frac{\hbar^2}{2m} \sum_{i=1}^N \nabla_i^2 + U(\{ {\bf r} \})  \label{H1}
\end{eqnarray}
where $\nabla_i^2$ is the Laplacian operator acting on the position ${\bf r}_i$ of particle $i$ and $U(\{ {\bf r} \})$ is the potential acting on all particle positions $\{ {\bf r} \} = \{{\bf r}_1,  \cdots , {\bf r}_N\}$. All the particles are considered to be identical and indistinguishable with mass $m$, and $\hbar = h/2\pi$ is the reduced Planck's constant. The quantum partition function for this system is \cite{McQuarrie2000}
\begin{equation}
Q_N = \sum_j e^{-\beta E_j}     \label{qQ}
\end{equation}
where $\beta = 1/k_BT$, $k_B$ is Boltzmann's constant and $E_j$ are the allowed energy states. The partition function can be written exactly in classical \emph{form}, following Kirkwood \cite{Kirkwood1933, Kirkwood1934} as described by McQuarrie \cite{McQuarrie2000}.
\begin{equation}
Q_N = \frac{1}{h^{3N}} \int \cdots \int e^{-\beta H} w(\{{\bf p}\},\{{\bf r}\},\beta) d\{{\bf p}\}d\{ {\bf r}\}  \label{Q1}
\end{equation}
where $\{ {\bf p} \}$ is the set of all particle momenta $\{{\bf p}_1,  \cdots , {\bf p}_N\}$, $d\{{\bf p}\}$ and $d\{ {\bf r}\}$ indicate integration over all momenta and positions, respectively, and $H$ is the classical Hamiltonian given by
\begin{equation}
H = \sum_{i=1}^N \frac{p_i^2}{2m} + U(\{ {\bf r} \})  . \label{H2}
\end{equation}
Note that equation (\ref{Q1}) is, for the moment, missing a factor $N!$ caused by ignoring symmetry considerations following McQuarrie \cite{McQuarrie2000}. For classical systems, the function $w(\{{\bf p}\},\{{\bf r}\},\beta) = 1$, but in general, for quantum systems, it must satisfy the relation
\begin{eqnarray}
e^{-\beta \mathcal{H}}e^{\frac{i}{\hbar} \sum_k {\bf p}_k \cdot {\bf r}_k} &=& e^{-\beta H} e^{\frac{i}{\hbar} \sum_k {\bf p}_k \cdot {\bf r}_k} w(\{{\bf p}\},\{{\bf r}\},\beta)  \nonumber \\
& \equiv & F(\{{\bf p}\},\{{\bf r}\},\beta)  . \label{McQ1}
\end{eqnarray}
This relation can be expressed instead as
\begin{eqnarray}
\frac{\partial F}{\partial \beta} &=& -\mathcal{H} F  \nonumber  \\
&=& \frac{\hbar^2}{2m}\sum_{i=1}^N \nabla_i^2 F - U(\{ {\bf r} \}) F \label{McQ2}
\end{eqnarray}
subject to the initial condition
\begin{equation}
F(\beta = 0) = e^{\frac{i}{\hbar} \sum_k {\bf p}_k \cdot {\bf r}_k}  .  \label{McQ3}
\end{equation}

The function $w(\{{\bf p}\},\{{\bf r}\},\beta)$ can be written in terms of a Hartree independent particle approximation, with a correlation correction, as follows.
\begin{equation}
w(\{{\bf p}\},\{{\bf r}\},\beta) = g_{xc} (\{{\bf r}\}, \beta) \prod_{i=1}^N \tilde{w}({\bf p}_i,{\bf r}_i,\beta) \label{w2}
\end{equation}
where $g_{xc} (\{{\bf r}\}, \beta)$ is taken to include not only quantum correlations, but also missing exchange effects required to enforce the indistinguishability of quantum particles, including the factor of $N!$. The $g_{xc}$ dependency on $\beta$ is not required, given the presence of $\beta$ in $\tilde{w}({\bf p}_i,{\bf r}_i,\beta) $, but it can be trivially included to allow the rephrasing of this exchange-correlation term in the more convenient form
\begin{equation}
g_{xc} (\{{\bf r}\}, \beta) \equiv e^{-\beta U_{xc}(\{{\bf r}\})} .  \label{Uxc}
\end{equation}
The partition function (\ref{Q1}) thus becomes
\begin{equation}
Q_N = \frac{1}{h^{3N}} \int \cdots \int e^{-\beta \tilde{H}} \prod_{i=1}^N \tilde{w}({\bf p}_i,{\bf r}_i,\beta) d\{{\bf p}\}d\{ {\bf r}\}  \label{Q2}
\end{equation}
where $\tilde{H} \equiv H + U_{xc}$.

There is much experience in the polymer SCFT community dealing with partition functions like (\ref{Q2}). Using Hubbard-Stratonovich transformations, the partition function can be rephrased, without approximation, into a form involving a density function $n({\bf r})$ and a conjugate chemical potential field $w({\bf r})$, expressed using Kac-Feynman path integrals \cite{Matsen2006, Qiu2006, Fredrickson2002, Schmid1998} . A mean field approximation then gives a partition function which, if classical correlations missing from the mean field are included in the exchange-correlation term, can be considered exact. The term ``exact'' is used throughout this paper in the context of DFT, that is, expressions are exact if a ``perfect'' exchange-correlation function is known. A free energy is readily obtained from the field theory transformed partition function through $F = -k_BT \ln Q_N$. The details of this process are widely available \cite{Matsen2006, Qiu2006, Fredrickson2002, Schmid1998}, but for completeness, a summary is provided in appendix \ref{derivation}.

From appendix \ref{derivation}, the SCFT equations are
\begin{eqnarray}
w({\bf r}) &=& \frac{\delta U[n]}{\delta n({\bf r})}  \label{w3} \\
n({\bf r}) &=& \frac{n_0}{Q} q({\bf r},{\bf r},\beta)    \label{n3}
\end{eqnarray}
where
\begin{equation}
Q = \frac{1}{V} \int d{\bf r} q({\bf r},{\bf r},\beta)   \label{Q3}
\end{equation}
and
\begin{equation}
\frac{\partial q({\bf r}_0,{\bf r},\beta)}{\partial \beta} = \frac{\hbar^2}{2m} \nabla^2 q({\bf r}_0,{\bf r},\beta) - w({\bf r}) q({\bf r}_0,{\bf r},\beta)  \label{diff3}
\end{equation}
subject to the initial conditions
\begin{equation}
q({\bf r}_0,{\bf r},0) = V \delta({\bf r}-{\bf r}_0)  .  \label{init3}
\end{equation}
A factor of the volume is absorbed into $q({\bf r},{\bf r},\beta) \equiv V \tilde{q}({\bf r},{\bf r},\beta)$ (see appendix \ref{derivation}), $n_0 \equiv N/V$ and the tilde from the appendix is dropped on $Q$. The function $q({\bf r}_0,{\bf r},\beta)$ may be interpreted as the unnormalized probability of a particle at a high classical temperature ($\beta = 0$) known to be at position ${\bf r}_0$, being found at position ${\bf r}$ when at a low quantum temperature ($\beta > 0$). The free energy is 
\begin{equation}
F[n,w] = -\frac{N}{\beta}\ln Q + U[n] - \int d{\bf r} w({\bf r}) n({\bf r})  .    \label{FE3}
\end{equation}
This set of equations is identical to the SCFT equations describing ring polymers \cite{Kim2012}, except in three ways. First, for polymers, the density of segments is summed (integrated) over all contour values whereas for the thermal trajectory of quantum particles, one is only interested in the density at a single temperature (no integral over contour in equation (\ref{n3})). Second, and more importantly, the potential $U[n]$ in (\ref{w3}) and (\ref{FE3}) is, of course, very different between polymers and quantum particles. For DFT applications, $U[n]$ will typically include an external potential (the ionic Coulomb potential), the electron-electron Coulomb interactions and an exchange-correlation functional. The third difference relates to the exchange-correlation functional, which is analogous to an equation of state in polymer SCFT, usually simple incompressibility. As presented here, $U[n]$ will also need to include the Pauli exclusion principle. All these potential terms are detailed in appendix \ref{sphB}. If the exchange-correlation term is assumed to fully enforce the exclusion principle, then the set of equations becomes the same as KS-DFT . This is shown in appendix \ref{kohnsham}. In the limit $h \rightarrow 0$, the free energy (\ref{FE3}) reduces to the classical DFT expression, as shown in appendix \ref{classicalDFT}.

Using the same algebra as in the classical limit (appendix \ref{classicalDFT}), the quantum free energy (\ref{FE3}) can be rephrased in terms of thermodynamic components \cite{Matsen1997} giving
\begin{equation}
F = \frac{1}{\beta} \int d{\bf r} n({\bf r}) \ln \left[ \frac{n({\bf r})}{n_0}\right] + U[n] - \frac{1}{\beta} \int d{\bf r} n({\bf r}) \left[\ln q({\bf r},{\bf r},\beta)  + \beta w({\bf r}) \right] . \label{FEcomp}
\end{equation}
The first term on the right hand side is the classical translational entropy of the quantum particles, and the second term includes all potential terms as discussed previously and in appendix \ref{sphB}. For polymer systems, the last term on the right hand side represents the polymer configurational entropy. For quantum particles, this corresponds to the non-interacting kinetic energy, in excess of the homogeneous. Since it is an excess quantity, it does not reduce to the Thomas-Fermi function, which is exact for homogeneous electron densities \cite{Carter2018}, but rather to zero in the uniform limit. This parallels the polymer case, where the polymer configurational entropy in SCFT becomes zero for uniform systems, even though there is certainly configurational entropy present. The homogeneous configurational entropy can be added, when needed, from known expressions. For quantum particles, the situation is the same, and as will be discussed shortly, the Thomas-Fermi energy is not needed for the systems studied here. For bosonic systems at zero temperature, the quantum non-interacting kinetic energy term of (\ref{FEcomp}) reduces to the ground state approximation for polymer configurational entropy as shown by Matsen \cite{Matsen2006}, which is the mathematically equivalent form to the von Weizs\"{a}cker functional. Expression (\ref{FEcomp}) is an \emph{exact} expression, assuming an exact input field $w({\bf r})$. For KS-DFT, the ``exact'' kinetic energy is limited by the inexactness of the exchange-correlation term. Here, the kinetic energy is additionally limited by the approximate Pauli exclusion principle, which is the common problem for all OF-DFT approaches \cite{Carter2018, Wang2000, Finzel2017}. There has been progress however in finding OF-DFT expressions that approach the accuracy of KS-DFT \cite{Levamaki2015}.

To obtain the equilibrium electron density, one self-consistently solves equations (\ref{w3})-(\ref{init3}) numerically. The computational limiting factor for doing this is solving the set of diffusion equations (\ref{diff3}), one for each spatial position ${\bf r}_0$. This is very computationally demanding, as discussed by Kim \cite{Kim2012}, but Matsen has observed that for ring polymers, this set of equations can be solved spectrally at the same numerical cost as solving a linear polymer with a \emph{single} diffusion equation \cite{Matsen2019}. This represents an enormous computational saving and makes possible the study of complicated three dimensional systems, as is routinely done in polymer SCFT \cite{Matsen2006, Qiu2006, Fredrickson2002, Schmid1998}. In the spectral method, one expands all spatially dependent functions in a superposition of orthonormal basis functions chosen to be eigenfunctions of the Laplacian operator and that encode the symmetry of the physical system. The method is well documented in the polymer SCFT community \cite{Matsen2006, Matsen2009} and, for completeness, it is summarized in appendix \ref{spectral}. The main result is that one must find the eigenvectors and eigenvalues of a single matrix (\ref{Aij}) which, through equation (\ref{specsolve3}), gives the propagators $q({\bf r}_0,{\bf r},\beta)$ that solve (\ref{diff3}). This is done once for every iteration towards self-consistency of the set of equations (\ref{w3})-(\ref{init3}). Compared to standard numerical approaches to DFT, this method requires the computational equivalent of finding the solution to a single diffusion equation each cycle compared to a single eigenvalue equation per cycle (OF-DFT) or a set of $N$ eigenvalue equations per cycle (KS-DFT). The method is particularly powerful in that the symmetry of the problem can be encoded in the basis set. Therefore, in polymer SCFT, it is the most computationally efficient method of solving the inhomogeneous density profiles for polymer self-assembly \cite{Matsen2009}, more so than either real space methods or pseudo-spectral methods that use fast Fourier transforms (FFTs) \cite{Rasmussen2002a, Rasmussen2002b}. FFTs are often used in OF-DFT, and it may be that the present spectral method could outperform FFT methods for systems with well-defined symmetry. Therefore, the best ultimate application for the present numerical approach to DFT might be to periodic solid state materials.

A simpler benchmark for the validity of the current set of equations (\ref{w3})-(\ref{init3}), in particular the governing diffusion equation (\ref{diff3}), is the ensemble average electron densities of atomic systems. This represents at the same time the simplest possible systems, as the ensemble average electron densities must have spherical symmetry (unlike orbitals) \cite{Cohen1965, Sagar1988} , while still being a rigorous test due to the highly inhomogeneous nature of the shell structure of atoms \cite{vonBarth2004}. A suitable basis set for spherically symmetric, ensemble average, electron densities are the zeroth order spherical Bessel functions which, when normalized, are given by
\begin{equation}
f_n(r) = \sqrt{\frac{2}{3}} \frac{R}{r} \sin \left(\frac{n\pi r}{R}\right)   \label{sphbessel}
\end{equation}
where $R$ is the radius of a finite spherical box, chosen large enough so that the electron densities go to zero on the boundary. In the limit of an infinite box, the overall electron density tends to zero, meaning that the Thomas-Fermi uniform electron density expression vanishes for this system, as previously mentioned \footnote{The $Z \rightarrow \infty$ limit of a neutral atom also approaches the Thomas-Fermi expression. Lee et al. \cite{Lee2009} and Cancio and Redd \cite{Cancio2017} have discussed how the inhomogeneity of the electron density gradually ``turns off'' in this limit, suggesting that the excess non-interacting quantum kinetic energy should approach zero in this case too, as expected.}. Spherical Bessel functions are not typically used for atomic and molecular calculations in DFT; Gaussian basis sets are the standard. One may consider trying to use Gaussians in the present context, but implementing the most scalable or computationally efficient basis set is not the objective of this paper. Rather, one would like to verify that the formalism reproduces expected results in the stringent, yet simple, atomic system. Spherical Bessel functions are very simple in that, unlike Gaussians, they are an orthonormal and complete set that are eigenfunctions of the Laplacian. The use of more complicated basis sets could obscure the results. Various quantities expanded in terms of the basis set (\ref{sphbessel}) are given in appendix \ref{sphB}. 

The equations (\ref{w3})-(\ref{init3}) are solved numerically and self-consistently using standard polymer SCFT algorithms, in particular, the spectral representation of appendix \ref{sphB}. The equations are applied to the set of atoms, hydrogen (H), helium (He), beryllium (Be), neon (Ne) and argon (Ar) in order to compare with the results of Finzel \cite{Finzel2015}. Picard iteration and Anderson mixing are used for convergence following Thompson \cite{Thompson2012} and self-consistent cycles are allowed to continue until the field coefficients stopped changing by less than one part in at most $10^{-8}$, although often to less than $10^{-9}$, according to the square of an L2-norm. In real space, this corresponds to the field changing by less than one part in at most $10^{-6}$ (often less than $10^{-7}$) according to the criterion of Finzel \cite{Finzel2015}.

\section{Results}

\begin{figure}
\includegraphics[width=1.0\textwidth]{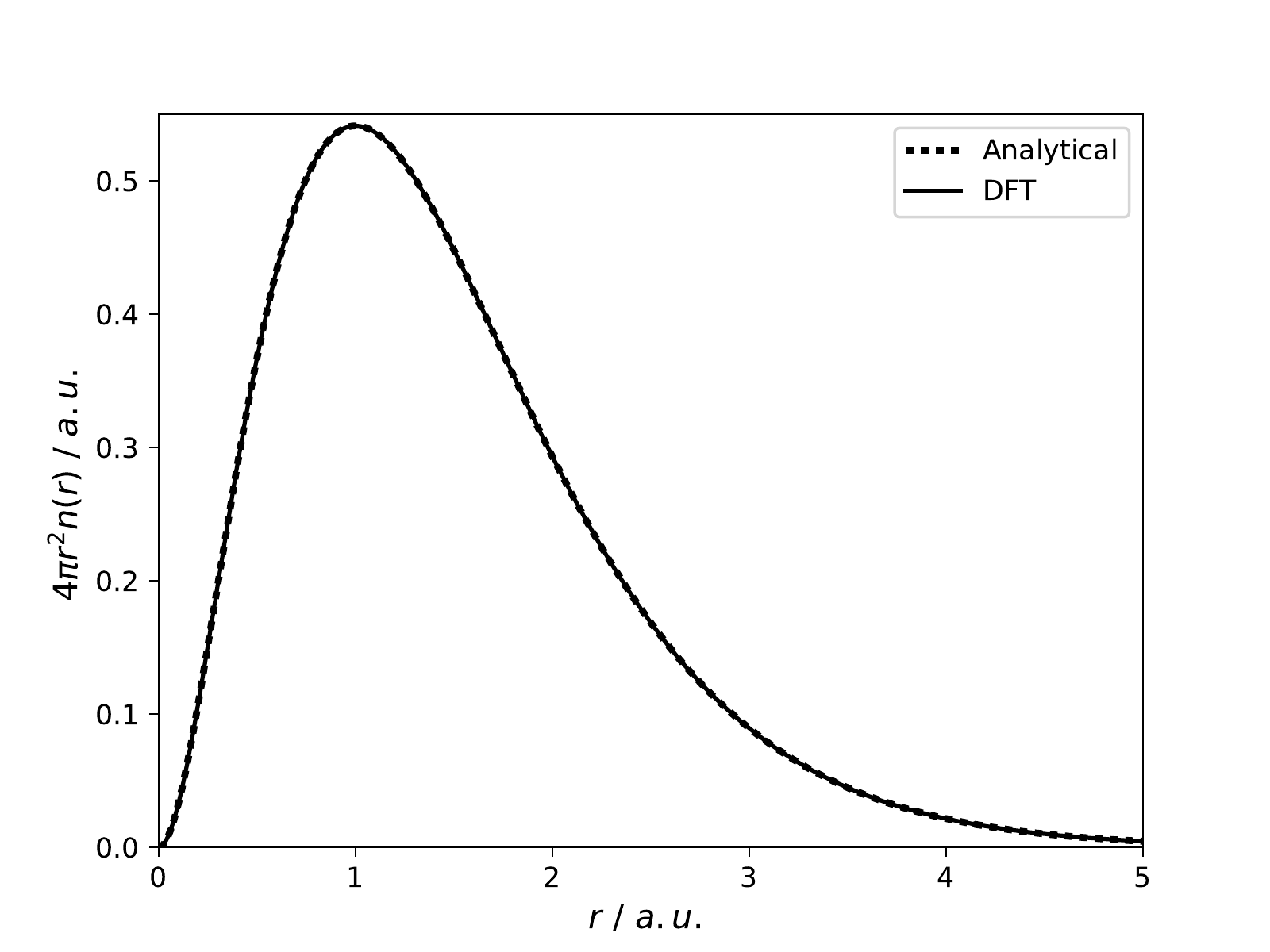}
\caption{Plot of radial electron density as a function of radius, in atomic units, for the hydrogen atom. The solid line is the numerical prediction and the dotted curve is the ground state analytical result.}
\label{fig:hydrogen}
\end{figure}

Analytical electron density results are known for the hydrogen atom, so the current formalism, and in particular equation (\ref{diff3}), can be given a basic test. The test is somewhat trivial, but exact, since there are no electron-electron, exchange or correlation interactions for the hydrogen system. Figure \ref{fig:hydrogen} shows the numerical results of equations (\ref{w3})-(\ref{init3}) together with the ground state analytical electron density curve, plotting the radial electron density against the atomic radius, both in atomic units. 
In figure \ref{fig:hydrogen}, and throughout the rest of this work, $\beta$ was taken sufficiently large so that the electron density and the corresponding free energy reached steady state. That is, the temperature was taken sufficiently low so that the ground state electron density was found. In principle, one could test temperature dependent results against analytical predictions, but this would require very high temperatures for a noticeable difference in the electron distribution. At such high temperatures, ionization would also be noticeable, making comparisons with the analytical results awkward. For this reason, the current work will focus solely on the ground state. Despite this, the finite temperature nature of the formalism means that it does not suffer some of the ambiguities of ground-state DFT \cite{Mermin1965, Argaman2002}. Instead of temperature dependence, one can check the effect of a finite size box on the hydrogen electron density, as these results are available \cite{Sommerfeld1938, Suryanarayana1976}. For box radii bigger than approximately 7 bohr, the energy of the hydrogen atom approaches the limit of -0.5 hartree, but as the radius of an enclosing box is reduced, the energy increases, passing through zero and becoming positive. In the present calculation, a zero energy is found for a box radius of 1.835 bohr, in perfect agreement with the prediction of Sommerfeld and Welker \cite{Sommerfeld1938}. For non-interacting electrons then, the current formalism, including equation (\ref{diff3}), seems to be a correct statistical mechanical description.  

For a less trivial test, the helium atom requires the Hartree electron-electron potential and an exchange-correlation function. In this work, the aim is not quantitative precision, but rather fidelity between the current approach and other DFT results. For this reason, the simplest possible local density approximation for exchange with no correlation is being used (LDAX) given by the functional \cite{Dirac1930, Finzel2015}
\begin{equation}
F_x[n] = -\frac{3}{4}\left(\frac{3}{\pi}\right)^{\frac{1}{3}}\int n({\bf r})^{\frac{4}{3}}d{\bf r} . \label{ldax}
\end{equation}
The density is now calculated self-consistently using (\ref{ldax}) and the Hartree potential given in appendix \ref{sphB}, with results shown in figure \ref{fig:helium}.
\begin{figure}
\includegraphics[width=1.0\textwidth]{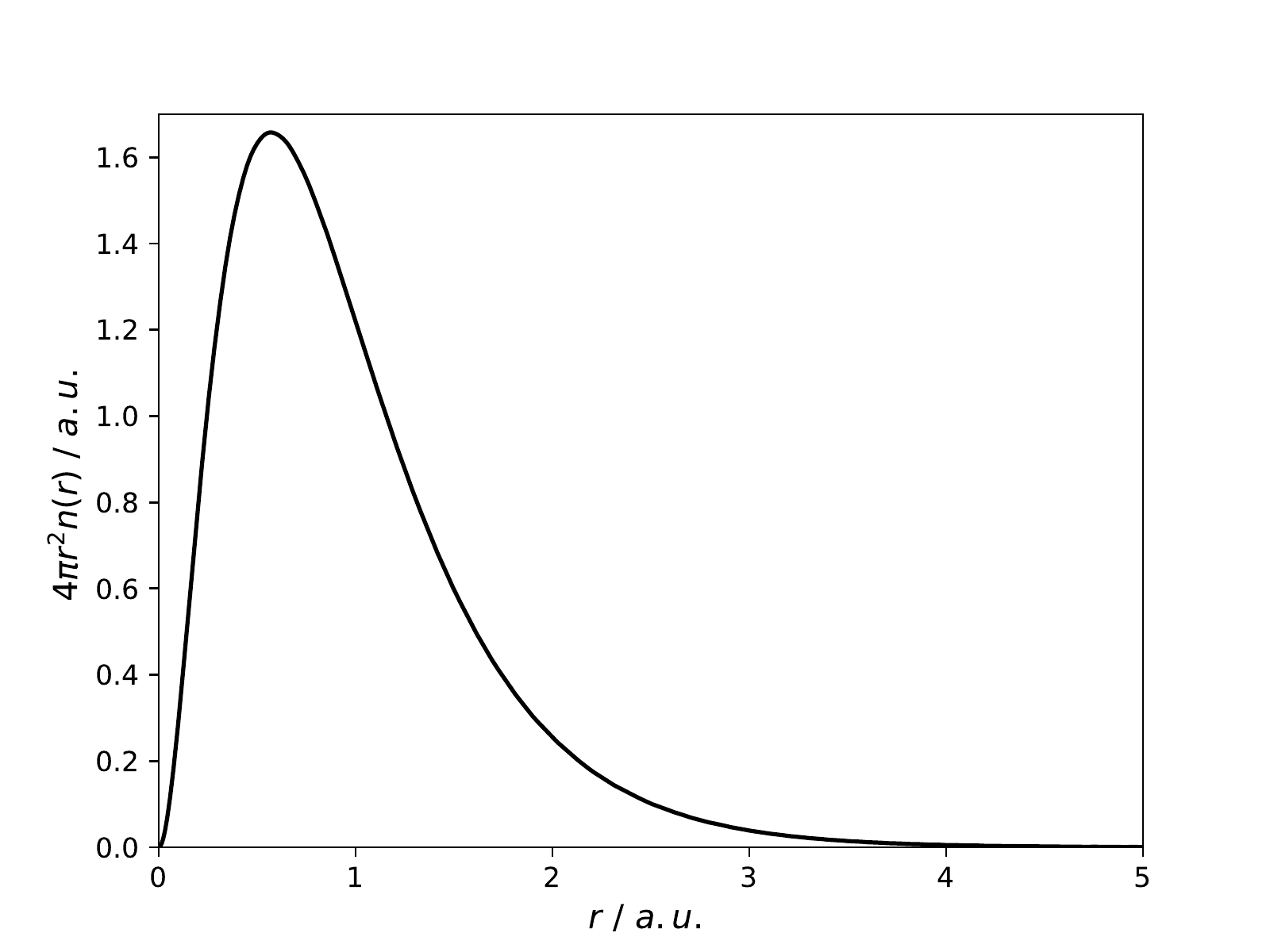}
\caption{Plot of radial electron density as a function of radius, in atomic units, for the helium atom.}
\label{fig:helium}
\end{figure}
This density profile is consistent with standard DFT predictions using the same LDAX approximation, as can be seen from a comparison with figure 1 of Finzel \cite{Finzel2015}.

\begin{figure}
\includegraphics[width=1.0\textwidth]{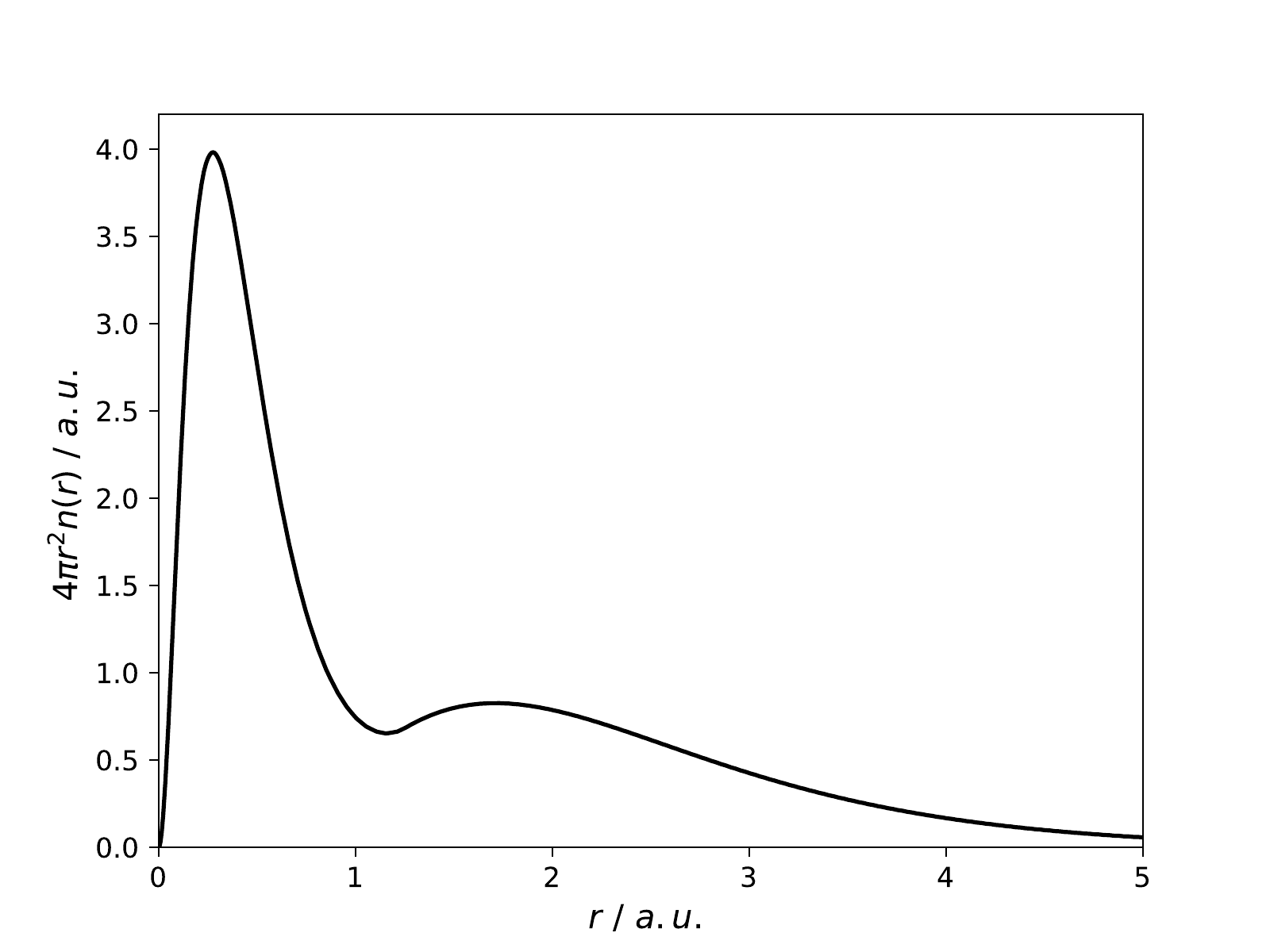}
\caption{Plot of radial electron density as a function of radius, in atomic units, for the beryllium atom.}
\label{fig:beryllium}
\end{figure}

Helium is also trivial in that it is essentially bosonic, that is, both electrons can exist in the ground state. Continuing comparisons with Finzel \cite{Finzel2015}, the density profile for beryllium is shown in figure \ref{fig:beryllium}. Here, a shell-structure-based (SSB)  potential has been added to the LDAX to enforce the Pauli exclusion principle. Such potentials have also been applied to systems more complicated than atoms, for example, solid state materials \cite{Finzel2015b}. For benchmarking purposes, the simplest possible SSB potential was used \cite{Finzel2015}.  
Comparing with figure 2 of reference \onlinecite{Finzel2015}, the electron profiles appear identical.

\begin{figure}
\includegraphics[width=1.0\textwidth]{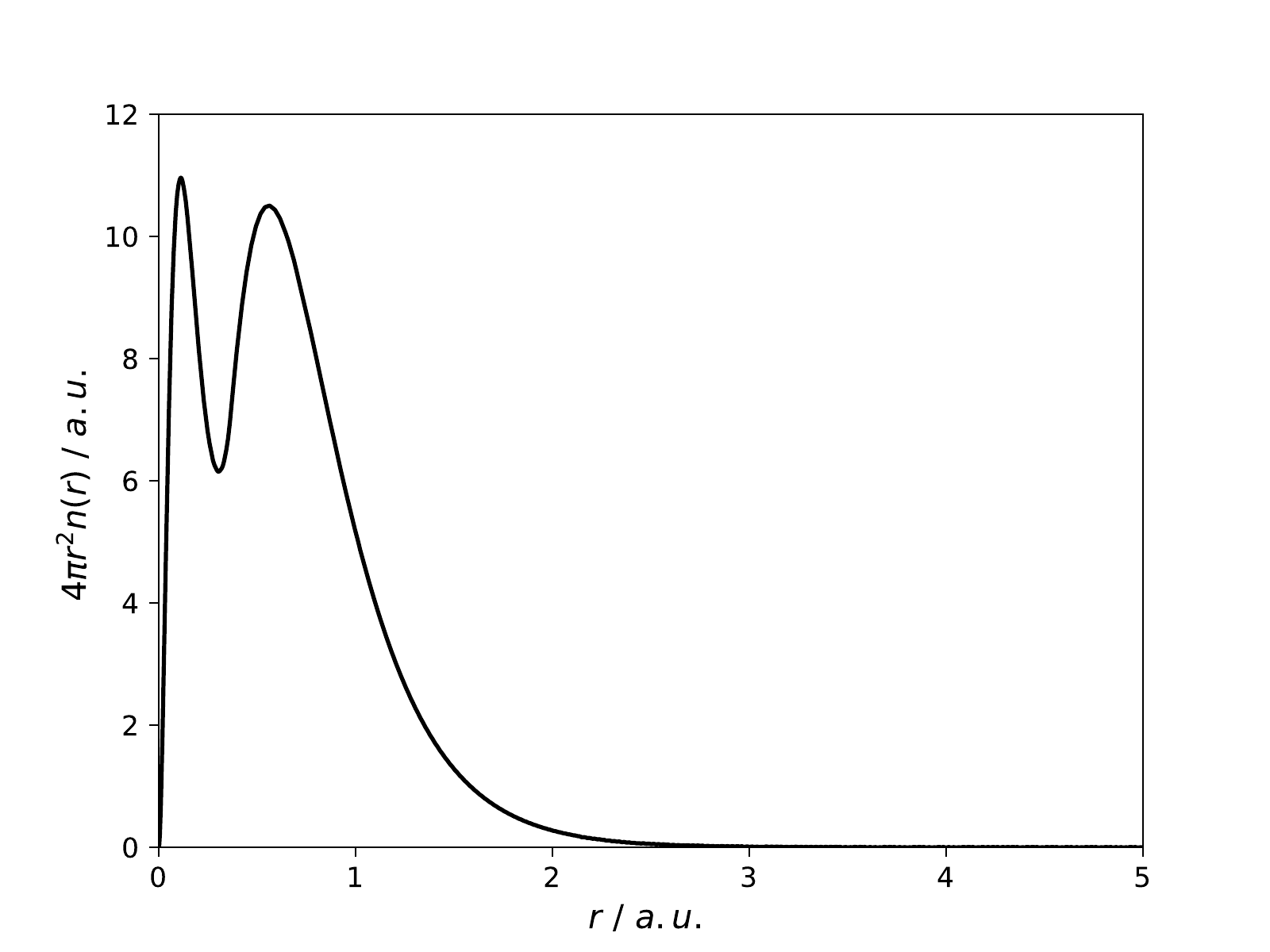}
\caption{Plot of radial electron density as a function of radius, in atomic units, for the neon atom.}
\label{fig:neon}
\end{figure}

\begin{figure}
\includegraphics[width=1.0\textwidth]{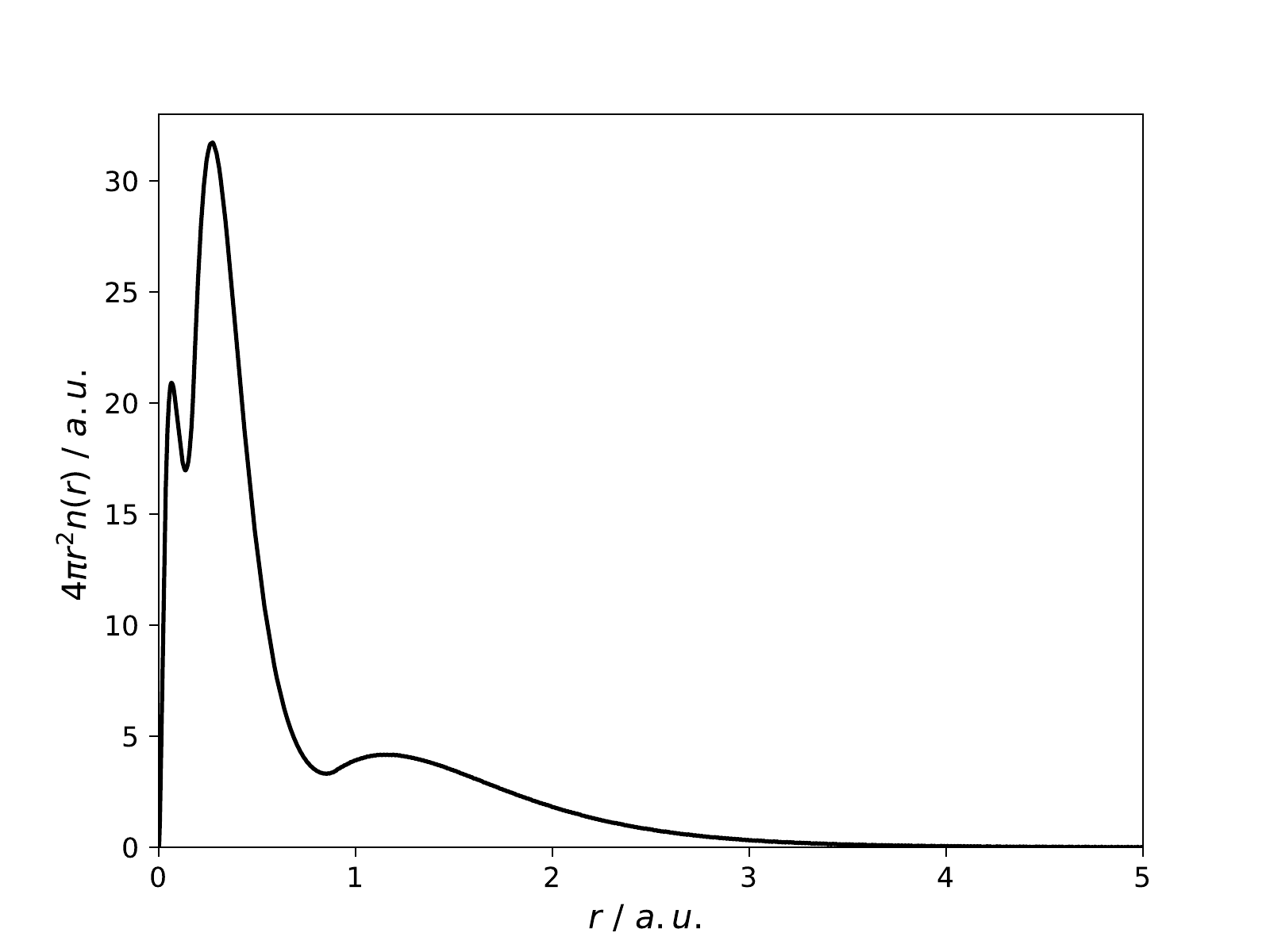}
\caption{Plot of radial electron density as a function of radius, in atomic units, for the argon atom.}
\label{fig:argon}
\end{figure}

Finzel also gives LDAX-SSB results for neon and argon (figures 3 and 4 of reference \onlinecite{Finzel2015}). The electron densities for these two atoms are calculated here and shown in figures \ref{fig:neon} and \ref{fig:argon}.
Again, the agreement is excellent. The Ne and Ar profiles are almost identical to Finzel's results, except for a slight difference in the tails. Most notably for argon, the present results show the distribution tail extending beyond 3 bohr, whereas for Finzel, the density is approximately zero by 2.5 bohr. Related to this, the smallest peak for argon is somewhat lower in this work than for Finzel. It was confirmed that these slight differences are not related to the temperature dependence of this formalism. Finzel notes that her numerical method provides ``fast and reliable convergence'' but that any deviation from the exact self-consistent solution ``manifests especially in the tail region'' \cite{Finzel2015}. Whatever the cause of the minor disagreement, it occurs for systems with higher numbers of electrons and so would not seem to be related to the validity of the single quantum particle governing equation (\ref{diff3}). 

\section{Discussion}

The derivation of DFT using the diffusion equation (\ref{diff3}) given in the Theory section and appendix \ref{derivation} starts from the partition function of quantum statistical mechanics, but finishes with equations that are essentially identical to SCFT ring polymers \cite{Kim2012}, except that instead of a polymer contour parameter $s$, which is a spatial parameterization variable embedded in real space, the contour for quantum particles is a thermal trajectory of the temperature parameter $\beta = 1/k_BT$, which is an independent variable. This justifies some elementary speculation on the relationship between the polymer and quantum particle pictures. 

In polymer SCFT, the diffusion equation (\ref{diff3}) gives the probability of finding a polymer segment $s$ at position ${\bf r}$ if the first segment $s=0$ is known to be at ${\bf r}_0$. The field $w({\bf r})$ in (\ref{diff3}) causes the polymer to deviate, or \emph{stretch}, away from the random walk it is known to have in polymer melts or theta solvent, and this stretching reduces the configurational entropy of the polymer, contributing to raising the free energy. By stretching to avoid the field $w({\bf r})$ however, the polymer reduces interaction energies, contributing to lowering the free energy. It is this competition, or frustration, between entropy (stretching) and energy, that determines the polymer behaviour in different environments. This mental picture may be applied to quantum particles, with the polymer stretching entropy interpreted as the quantum non-interacting kinetic energy. This follows because the DFT results presented in the Theory section could have been derived using \emph{classical} statistical mechanics on an extended, polymer-like object, in a four dimensional thermal-space, rather than using quantum statistical mechanics on a point-like particle in three dimensional space. The idea of using temperature as a dimension (imaginary time) is not new \cite{Matsubara1955, Das1993}; here, the perspective is suggested by the equivalence of the quantum, point-like, derivation to the known, classical, polymer-like, results. 

For this speculative picture to be valid for non-relativistic quantum mechanics (QM), one needs to follow authors like Ballentine \cite{Ballentine2003}, Bransden and Joachain \cite{Bransden2000}, or Aharonov \textit{et al.} \cite{Aharanov2017} who recommend an ensemble interpretation to quantum phenomena. In principle, all results of QM would be obtained by calculating SCFT on polymer-like objects using classical statistical mechanics in higher dimensions. Many mathematical postulates of QM would no longer be necessary, being replaced by the mathematics of classical statistical mechanics. The wave function postulate would be replaced by the assumption that quantum particles are polymer-like trajectories embedded in a four dimensional thermal-space. Again, this is not novel, in that it is known that a $(D+1)$ dimensional quantum field theory can be replaced by a $D$ dimensional quantum statistical mechanical system \cite{Das1993}, although here, it is being suggested that a $(D+1)$ dimensional classical statistical field theory can replace $D$ dimensional quantum mechanics. Replacing $\beta$ in the diffusion equation (\ref{diff3}) with the complex time $it/\hbar$ (Wick rotation) and integrating over ${\bf r}_0$ gives the time-dependent Schr\"{o}dinger equation with a static potential. In fact, both the Schr\"{o}dinger equation and the wave function map mathematically onto SCFT, as pointed out by Matsen \cite{Matsen2006}. The wave function could continue to be viewed operationally as a mathematical tool without physical significance on its own, as in the Copenhagen interpretation, but now it would actually be the projection of the four dimensional thermal-space system onto three dimensional space. By using the wave function, one would not always have to perform numerically involved, four dimensional (or five dimensional if time dependent) classical statistical mechanics to get any results. A ``thought calculation'' using the double-slit experiment can demonstrate that the 4D polymeric view of quantum particles will result in a three dimensional wave behaviour. 

In the ensemble picture of QM, a succession of independent quantum particles shot through a double slit at a screen can be viewed together as an incident group. Upon hitting the double slit, classical arguments would predict that most of the incident group would be blocked, and only two localized spots on the screen directly aligned with the slits would be observed. The governing equation of the SCFT approach is the diffusion equation (\ref{diff3}) which gives the probability that a particle will be found at position ${\bf r}$ at a temperature $\beta$, if it is known with certainty to be at position ${\bf r}_0$ at a high, classical, temperature. For large enough $\beta$ (low enough temperature), the probability of finding the particle at ${\bf r}$ starts to diffuse, or de-localize. For the incident ensemble group, the spherical diffusions will sum to give a plane wave of probability. Upon hitting the screen however, only two point sources will continue to diffusively expand on the other side, giving two spherical probability waves which will interfere in the usual way. This can be pictured as the four dimensional ``polymers'' stretching around the slits to give non-zero probabilities even in regions not directly aligned with the slits. Thus spatial 3D probabilities can be viewed in terms of waves, or wave functions, in agreement with standard QM. This is essentially a realist, ``pilot-wave'', perspective, but with polymeric thermal pilot-trajectories, as suggested from the QM statistical mechanics derivation, instead of an actual pilot wave. As such, there are no issues with interpreting the ``collapse'' of the wave function, and the quantum particles can be viewed as having an objective existence outside of observation. Also, since the quantum particles are viewed as extended, non-local objects, the Bell theorem proscription on hidden variables may not apply \cite{Bell1966}. 

Of course, a classical statistical mechanical view with hidden variables begs the question of what the hidden variables are. In typical classical statistical mechanics, thermal Brownian motion accounts for a mechanism in which constituents explore the phase space over the ensemble, or from the ergodic hypothesis, over time. No explanation is given in the current derivation of why a non-local thermal diffusive thread described by (\ref{diff3}) for a single quantum particle should randomly explore configurations in a vacuum. Nor is there any first principles derivation of the origin of the Pauli exclusion principle, since the overall derivation is bosonic, in common with other OF-DFTs \cite{Wang2000, March2010, Karasiev2014, Finzel2017, Carter2018, Finzel2017, Finzel2016b, Finzel2015} . Of course, inherent randomness and Pauli exclusion are not addressed within non-relativistic QM theory either, so the current perspective neither adds nor detracts from standard QM. It does, perhaps, provide an alternative viewpoint and platform from which such questions might be addressed in the future.

\section{Conclusions}

A derivation of quantum density functional theory has been given that is mathematically equivalent to a classical statistical mechanical derivation of ring polymers in a four dimensional thermal-space. From this perspective, quantum mechanics is viewed in the ensemble interpretation as a fundamentally thermal theory, and one that predicts wave-like behaviour in three spatial dimensions. The wave function can be viewed as a mathematical tool describing the projection of the classical statistical mechanical results in four dimensional temperature-space to three dimensional space. 

As a first test of the governing equations derived here, the ensemble average electron densities of isolated H, He, Be, Ne and Ar atoms were calculated using an exchange-only local density approximation and a shell-structured-based Pauli potential. A spectral method, which reduces the set of modified diffusion equations to the numerical equivalent of a single equation, was used. Since this equation does not scale with the number of electrons, all-electron calculations were performed without pseudo-potentials. Also, to avoid confounding factors, an orthonormal, complete, spherical Bessel basis set was used rather than Gaussians. Agreement with the results of Finzel are excellent \cite{Finzel2015}. 

The atomic system is rigorous test, because of the highly inhomogeneous shell structure, and yet it is also a simple benchmark, due to its symmetry. The numerical approach may be particularly well suited however to periodic solid state materials, where a Fourier basis set can be used that incorporates the symmetry of the crystal. This follows the example of the use of the spectral method in polymer SCFT, where the spectral method has been shown to be the most numerically efficient technique, significantly better than real space and fast Fourier transform pseudo-spectral methods in most cases \cite{Matsen2009}. For molecular systems, the use of Gaussian basis sets would be more practical and their suitability in the present formalism could be investigated. Spin could also be incorporated by calculating two densities, one for each spin state. The current application deliberately uses a simple approximation for the Pauli potential, but alternative SSB Pauli potentials are possible \cite{Wang1982, Sagar1988, Schmider1992, Tsirelson2013, Finzel2016, Finzel2016b} as are improved approximations for the exchange-correlation functional \cite{Parr1989, vonBarth2004, Becke2014, Jones2015}.

\section{Acknowledgements}

The author is grateful to M. W. Matsen for pointing out that a set of ring polymer diffusion equations can be solved at the computational cost of a single diffusion equation by using a spectral representation, and to an anonymous reviewer for mentioning the $Z \rightarrow \infty$ neutral atom Thomas-Fermi limit. The author thanks J. Z. Y. Chen and M. W. Matsen for commenting on the manuscript prior to publication. This research was financially supported by the Natural Sciences and Engineering Research Council of Canada (NSERC).

\appendix

\section{SCFT Derivation}  \label{derivation}

Following von Barth \cite{vonBarth2004}, a density operator can be defined as
\begin{equation}
\hat{n}({\bf r}) = \sum_{i=1}^N \delta ({\bf r} - {\bf r}_i)     \label{density1}
\end{equation}
so that the Hamiltonian, including $U_{xc}$, can be written as
\begin{equation}
\tilde{H} = \sum_{i=1}^N \frac{p_i^2}{2m} + U[\hat{n}]   \label{H3}
\end{equation}
where $U[\hat{n}]$ is the functional expression of the potential in terms of the density operator $\hat{n}({\bf r})$. The partition function is then
\begin{equation}
Q_N = \frac{1}{h^{3N}} \int \cdots \int e^{-\beta U[\hat{n}]} \prod_{i=1}^N e^{\frac{-\beta p_i^2}{2m}} \tilde{w}({\bf p}_i,{\bf r}_i,\beta) d\{{\bf p}\}d\{ {\bf r}\}  . \label{Q9}
\end{equation}
The definition of the functional Dirac delta is \cite{Matsen1994}
\begin{equation}
\int \mathcal{D}\mathcal{N} \delta[\mathcal{N} - \hat{n}] G[\mathcal{N}] = G[\hat{n}]  \label{Dirac1}
\end{equation}
where $G$ is an arbitrary functional. This can be inserted into (\ref{Q9}) to give
\begin{equation}
Q_N = \frac{1}{h^{3N}} \int \mathcal{D} \mathcal{N} \delta[\mathcal{N} - \hat{n}] e^{-\beta U[\mathcal{N}]} \int \cdots \int  \prod_{i=1}^N e^{\frac{-\beta p_i^2}{2m}} \tilde{w}({\bf p}_i,{\bf r}_i,\beta) d\{{\bf p}\}d\{ {\bf r}\}  . \label{Q10}
\end{equation}
Using a particular form of functional Dirac delta \cite{Matsen1994},
\begin{equation}
\delta[\mathcal{N} - \hat{n}] = \int_{-i \infty}^{i\infty} \mathcal{D}W \exp\left\{\beta \int d{\bf r}^\prime W({\bf r}^\prime) \left[ \mathcal{N}({\bf r}^\prime) - \hat{n}({\bf r}^\prime)\right] \right\}   \label{Dirac2}
\end{equation}
allows (\ref{Q10}) to be written as
\begin{equation}
Q_N = \frac{1}{h^{3N}} \int \mathcal{D}\mathcal{N} \mathcal{D} W  e^{-\beta U[\mathcal{N}] + \beta \int d{\bf r}^\prime W({\bf r}^\prime) [ \mathcal{N}({\bf r}^\prime) - \hat{n}({\bf r}^\prime)]} \int \cdots \int\prod_{i=1}^N e^{\frac{-\beta p_i^2}{2m}} \tilde{w}({\bf p}_i,{\bf r}_i,\beta) d\{{\bf p}\}d\{ {\bf r}\}  . \label{Q11}
\end{equation}
Noting, from equation (\ref{density1}), that 
\begin{equation}
-\beta \int d{\bf r}^\prime W({\bf r}^\prime)\hat{n}({\bf r}^\prime) = -\beta \int d{\bf r}^\prime W({\bf r}^\prime)\sum_{i=1}^N \delta ({\bf r} - {\bf r}_i) = -\beta \sum_{i=1}^N W({\bf r}_i)  \label{w4}
\end{equation}
allows the partition function (\ref{Q11}) to be rewritten as
\begin{equation}
Q_N = \frac{1}{h^{3N}} \int \mathcal{D}\mathcal{N} \mathcal{D} W  e^{-\beta U[\mathcal{N}] + \beta \int d{\bf r}^\prime W({\bf r}^\prime) \mathcal{N}({\bf r}^\prime)} \int \cdots \int\prod_{i=1}^N e^{-\beta H_{{\rm eff}}({\bf p}_i,{\bf r}_i)} \tilde{w}({\bf p}_i,{\bf r}_i,\beta) d\{{\bf p}\}d\{ {\bf r}\}   \label{Q12}
\end{equation}
where
\begin{equation}
H_{{\rm eff}}({\bf p},{\bf r}) = \frac{p^2}{2m} + W({\bf r}) .  \label{Heff}
\end{equation}
A single particle partition function may be defined as
\begin{equation}
\tilde{Q} = \frac{1}{h^3} \int\int d{\bf p} d{\bf r} e^{-\beta H_{{\rm eff}}({\bf p},{\bf r})} \tilde{w} ({\bf p},{\bf r},\beta)  \label{Qtilde1}
\end{equation}
which has the same form as the original partition function (\ref{Q1}) except that it is for a single particle subject to the effective Hamiltonian (\ref{Heff}). It thus obeys the same McQuarrie relation (\ref{McQ1}), namely,
\begin{eqnarray}
e^{-\beta \mathcal{H}_{\rm KS}}e^{\frac{i}{\hbar} {\bf p} \cdot {\bf r}} &=& e^{-\beta H_{\rm eff}} e^{\frac{i}{\hbar} {\bf p} \cdot {\bf r}} w({\bf p},{\bf r},\beta)  \nonumber \\
& \equiv & \tilde{q}({\bf p},{\bf r},\beta)   \label{McQ4}
\end{eqnarray}
where 
\begin{equation}
\mathcal{H}_{{\rm KS}} = -\frac{\hbar^2}{2m} \nabla^2 + W({\bf r})   .\label{Heff2}
\end{equation}
The subscript ``KS'' on (\ref{Heff2}) indicates that this is the same operator as in the Kohn-Sham equations. Equation (\ref{McQ4}) obeys the relation
\begin{equation}
\frac{\partial \tilde{q}({\bf p},{\bf r},\beta)}{\partial \beta} = \frac{\hbar^2}{2m} \nabla^2 \tilde{q}({\bf p},{\bf r},\beta) - W({\bf r}) \tilde{q}({\bf p},{\bf r},\beta)  \label{diff1}
\end{equation}
subject to the initial conditions
\begin{equation}
\tilde{q}({\bf p},{\bf r},0) = e^{\frac{1}{\hbar}{\bf p}\cdot{\bf r}} .  \label{init1}
\end{equation}
The single particle partition function $\tilde{Q}$ given by (\ref{Qtilde1}) can therefore be rewritten using (\ref{McQ4}) as 
\begin{equation}
\tilde{Q} = \frac{1}{h^3} \int\int d{\bf p} d{\bf r}  \tilde{q}({\bf p},{\bf r},\beta) \tilde{q}^*({\bf p},{\bf r},0) \label{Qtilde2}
\end{equation}
and the full partition function (\ref{Q12}) becomes
\begin{equation}
Q_N = \int \mathcal{D}\mathcal{N}\mathcal{D}W \tilde{Q}^N e^{-\beta U[\mathcal{N}] + \beta \int d{\bf r} W({\bf r}) \mathcal{N}({\bf r})} .  \label{Q7}
\end{equation}
The expression (\ref{Q7}) is formally exact, to the extent that a ``perfect'' exchange-correlation functional is known. A saddle function approximation resulting from the first term of a functional Taylor series expansion of (\ref{Q7}) \cite{Das1993} gives the partition function
\begin{equation}
Q_N = \tilde{Q}^N e^{-\beta U[n] + \beta \int d{\bf r} w({\bf r}) n({\bf r})}   \label{Q8}
\end{equation}
which may also be considered exact if classical correlations ignored in the saddle function approximation are included in the exchange-correlation potential. In (\ref{Q8}), $w({\bf r})$ and $n({\bf r})$ are the mean field values of the field and quantum particle density, respectively, about which the Taylor series is expanded \cite{Das1993}. The free energy is readily obtained through $F = -k_BT \ln Q_N$ giving
\begin{equation}
\frac{F[n,w]}{k_BT} = -N\ln \tilde{Q} + \beta U[n] - \beta \int d{\bf r} w({\bf r}) n({\bf r})  .\label{FE1}
\end{equation}
Again, this free energy is exact to the extent that the exchange-correlation potential is exact. 

The SCFT equations can be found by varying (\ref{FE1}) with respect to $n({\bf r})$ and $w({\bf r})$. This gives the pair of equations
\begin{eqnarray}
w({\bf r}) &=& \frac{\delta U[n]}{\delta n({\bf r})}  \label{w1} \\
n({\bf r}) &=& -\frac{N}{\tilde{Q}\beta}\frac{\delta \tilde{Q}}{\delta w({\bf r})}    \label{n1}
\end{eqnarray}
where it should be noticed that $n({\bf r})$ is also a function of the temperature parameter $\beta$, as is $\tilde{Q}$. This set of equation would be solved once the potentials in $U[n]$ are specified --- see appendix \ref{sphB}. It is easier to solve them if one rephrases the diffusion equation (\ref{diff1}) by Fourier transforming over the ${\bf p}$ coordinates to give
\begin{equation}
\frac{\partial \tilde{q}({\bf r}_0,{\bf r},\beta)}{\partial \beta} = \frac{\hbar^2}{2m} \nabla^2 \tilde{q}({\bf r}_0,{\bf r},\beta) - w({\bf r}) \tilde{q}({\bf r}_0,{\bf r},\beta)  \label{diff2}
\end{equation}
subject to the initial conditions
\begin{equation}
\tilde{q}({\bf r}_0,{\bf r},0) = \delta({\bf r}-{\bf r}_0)  .  \label{init2}
\end{equation}
Equation (\ref{diff2}) is analogous to the equation of motion for a single particle quantum propagator \cite{Fredrickson2006}. Performing the Fourier transform on the right hand side of (\ref{Qtilde2}) gives
\begin{equation}
\tilde{Q} = \int d{\bf r} \tilde{q}({\bf r},{\bf r},\beta) .  \label{Qtilde3}
\end{equation}
In the above equations, the Fourier transform of $\tilde{q}({\bf p},{\bf r},\beta)$ with respect to ${\bf p}$ is expressed as $\tilde{q}({\bf r}_0,{\bf r},\beta)$. The functional derivative in (\ref{n1}) can be performed giving
\begin{equation}
\frac{\delta \tilde{Q}}{\delta w({\bf r}^\prime)} = \int d{\bf r} \frac{\delta \tilde{q} ({\bf r},{\bf r},\beta)}{\delta w({\bf r}^\prime)}  .\label{der1}
\end{equation}
Equations (\ref{diff2}) and (\ref{init2}) are mathematically identical to the governing equations for a polymer \cite{Edwards1965, Helfand1975}. Therefore the formal Kac-Feynman solution is the same \cite{Matsen2006, Matsen1994, Helfand1975}.
\begin{equation}
\tilde{q}({\bf r}_0,{\bf r},\beta) = \mathcal{N} \int_{{\bf r}_0}^{{\bf r}} \mathcal{D} {\bf r} P\left[{\bf r};0,\beta\right] e^{-\int_0^\beta d\tau w({\bf r}(\tau))}  \label{Kac}
\end{equation}
where
\begin{equation}
P\left[{\bf r}^\prime;\tau_1,\tau_2 \right] \propto \exp\left[-\frac{m}{2\hbar^2}\int_{\tau_1}^{\tau_2} d\tau \left| \frac{d{\bf r}^\prime (\tau)}{d\tau}\right|^2\right]  \label{prob}
\end{equation}
and $\mathcal{N}$ is a normalization factor. Equations (\ref{Kac}) and (\ref{prob}) describe a ``thermal trajectory'' of a quantum particle. Using (\ref{diff2}), the functional derivative in (\ref{der1}) can be performed.
\begin{equation}
\frac{\delta \tilde{Q}}{\delta w({\bf r}^\prime)} = -\beta \tilde{q}({\bf r}^\prime,{\bf r},\beta)   \label{der2}
\end{equation}
and so the density (\ref{n1}) becomes
\begin{equation}
n({\bf r}) = \frac{N}{\tilde{Q}} \tilde{q}({\bf r},{\bf r},\beta)  .  \label{n2}
\end{equation}

\section{Kohn-Sham Equivalence}  \label{kohnsham}

The operator on the right hand side of the diffusion equation (\ref{diff3}) is
\begin{equation}
H_{KS} = \frac{\hbar^2}{2m} \nabla^2  - w({\bf r})  \label{Hks}
\end{equation}
where the subscript ``KS'' indicates that this is the same form as the operator in the Kohn-Sham equations. Let $\phi_i({\bf r})$ and $\varepsilon_i$ be the eigenfunctions and eigenvalues, respectively, of the operator (\ref{Hks}). The eigenvalue equations for $H_{KS}$ is
\begin{equation}
H_{KS} \phi_i({\bf r}) = \varepsilon_i \phi_i({\bf r})  \label{KS}
\end{equation}
which are the Kohn-Sham equations. The eigenfunctions can be chosen to be orthonormal according to
\begin{equation}
\frac{1}{V} \int d{\bf r} \phi_i({\bf r}) \phi_j^*({\bf r}) = \delta_{ij}    \label{ortho1}
\end{equation}
where $\delta_{ij}$ is the Kronecker delta. The functions $q({\bf r}_0,{\bf r},\beta)$ can be expanded in a basis set of the eigenfunctions of $H_{KS}$ to give
\begin{equation}
q({\bf r}_0,{\bf r},\beta) = \sum_{i=1}^\infty q_i({\bf r}_0,\beta) \phi_i({\bf r})  \label{expand1}
\end{equation}
and this can be substituted into the diffusion equation (\ref{diff3}). From the derivation of Matsen \cite{Matsen2006}, one finds
\begin{equation}
q({\bf r}_0,{\bf r},\beta) = \sum_{i=1}^\infty c_i({\bf r}_0) e^{\varepsilon_i \beta} \phi_i({\bf r})  \label{q1}
\end{equation}
where
\begin{eqnarray}
c_i({\bf r}_0) &=& q_i({\bf r}_0,0)  \nonumber \\
&=& \frac{1}{V} \int d{\bf r} \phi_i^*({\bf r}) q({\bf r}_0,{\bf r},0) \nonumber \\
&=&  \int d{\bf r} \phi_i^*({\bf r}) \delta({\bf r}-{\bf r}_0)  \nonumber \\
&=&  \phi_i^*({\bf r}_0)   \label{ci}
\end{eqnarray}
using (\ref{init3}). Therefore, (\ref{q1}) becomes
\begin{equation}
q({\bf r}_0,{\bf r},\beta) = \sum_{i=1}^\infty e^{\varepsilon_i \beta}\phi_i^*({\bf r}_0)  \phi_i({\bf r})  \label{q2}
\end{equation}
which is analogous to the eigenfunction representation of a quantum propagator \cite{Fredrickson2006}. From equation (\ref{n3}), the density becomes
\begin{eqnarray}
n({\bf r}) &=& \frac{n_0}{Q} q({\bf r},{\bf r},\beta) \nonumber  \\
&=& \frac{n_0}{Q}  \sum_{i=1}^\infty e^{\varepsilon_i \beta}  \left| \phi_i({\bf r}) \right|^2   \label{n4}
\end{eqnarray}
Similarly, the single particle partition function $Q$ from (\ref{Q3}) becomes
\begin{eqnarray}
Q &=& \frac{1}{V} \int d{\bf r} q({\bf r},{\bf r},\beta)  \nonumber \\ 
&=& \frac{1}{V} \sum_{i=1}^\infty e^{\varepsilon_i \beta} \int d{\bf r} \left| \phi_i({\bf r}) \right|^2  \nonumber \\
&=& \sum_{i=1}^\infty e^{\varepsilon_i \beta}  \label{Q4}
\end{eqnarray}
using orthonormality (\ref{ortho1}). Therefore the density is
\begin{equation}
n({\bf r}) = n_0 \frac{\sum_{i=1}^\infty e^{\varepsilon_i \beta} \left| \phi_i({\bf r}) \right|^2}{\sum_{i=1}^\infty e^{\varepsilon_i \beta}} .  \label{n5}
\end{equation}

The probability that a system is in a state $R$ is given by the canonical distribution \cite{Rief1965}
\begin{equation}
P_R = \frac{e^{-\beta E_R}}{\sum_{R^\prime} e^{-\beta E_{R^\prime}}}   \label{pr}
\end{equation}
where $E_R$ is the energy of state $R$. The average occupancy $\bar{n}_i$ of the state $i$ will be \cite{Rief1965}
\begin{equation}
\bar{n}_i = \sum_R n_i P_R   .  \label{occ}
\end{equation}
Following the Rief derivation of Fermi-Dirac statistics in the canonical ensemble \cite{Rief1965}, if the Pauli exclusion principle is enforced, comparing (\ref{n5}) with (\ref{pr}) and (\ref{occ}), the density will be
\begin{equation}
n({\bf r}) = \frac{1}{V}\sum_{i=1}^\infty f(\varepsilon_i - \mu) \left| \phi_i \right|^2  \label{n6}
\end{equation}
in agreement with Kohn and Sham \cite{Kohn1965}, where $f(\varepsilon_i - \mu)$ is the Fermi-Dirac distribution and $\mu$ is the chemical potential. For $T \rightarrow 0$, this becomes
\begin{equation}
n({\bf r}) = \frac{1}{V} \sum_{i=1}^\infty \left| \phi_i \right|^2  \label{n7}
\end{equation}
which is the standard formula for the density in KS-DFT \cite{Kohn1965} apart from the factor $1/V$. (This factor arises because the convention of Matsen \cite{Matsen2006} is being followed here in which $1/V$ is included in the orthogonality definition (\ref{ortho1}).) Note that for this equivalence with Kohn-Sham to be valid, the field $w({\bf r})$ would have to rigorously enforce the Pauli exclusion principle. Although this should be in principle true --- the exclusion principle arises due to the exchange symmetry of fermions, and so should be encoded in the exchange-correlation potential ---  in KS-DFT, the exclusion is put in ``by hand'' through the sum in (\ref{n7}) and the equivalent sum in the kinetic energy term. Thus typical exchange-correlation functionals do not enforce the Pauli principle completely. The formalism presented in this paper is therefore, operationally, an OF-DFT in that the Pauli exclusion principle needs to be added to the exchange-correlation functional, as discussed by Finzel and others \cite{Carter2018, Finzel2017, Finzel2016b, Finzel2015, Wang2000}.

\section{Classical DFT Limit}  \label{classicalDFT}

It can be shown that the set of equations (\ref{w3})-(\ref{FE3}) become equivalent to classical DFT for classical systems. In the classical limit, $h \rightarrow 0$, the Laplacian term in the diffusion equation (\ref{diff3}) disappears, giving 
\begin{equation}
\frac{\partial q({\bf r}_0,{\bf r},\beta)}{\partial \beta} = - w({\bf r}) q({\bf r}_0,{\bf r},\beta)  \label{diff4}
\end{equation}
subject to the initial condition (\ref{init3}). Equation (\ref{diff4}) can be solved analytically to give
\begin{equation}
q({\bf r}_0,{\bf r},\beta) = V \delta({\bf r}-{\bf r}_0) e^{-\beta w({\bf r})}    \label{qclass}
\end{equation}
and
\begin{equation}
q({\bf r},{\bf r},\beta) = V \delta(0) e^{-\beta w({\bf r})}   . \label{qclass2}
\end{equation}
From (\ref{Q3}), the single particle partition function becomes
\begin{eqnarray}
Q &=& \frac{1}{V} \int d{\bf r} q({\bf r},{\bf r},\beta)   \nonumber \\
&=&  \delta(0) \int d{\bf r} e^{-\beta w({\bf r})} . \label{Qclass}  
\end{eqnarray}
Therefore, from (\ref{n3}), the density becomes
\begin{eqnarray}
n({\bf r}) &=& \frac{n_0}{Q} q({\bf r},{\bf r},\beta)    \nonumber  \\
&=&  N \frac{e^{-\beta w({\bf r})}}{\int d{\bf r} e^{-\beta w({\bf r})}} . \label{nclass}
\end{eqnarray}
This can be rearranged as 
\begin{equation}
w({\bf r}) =- \frac{1}{\beta} \ln \left[\frac{Qn({\bf r})}{n_0}\right]  \label{wclass}
\end{equation}
which can be used in the free energy expression (\ref{FE3}) to give
\begin{equation}
\frac{F[n]}{k_BT} = \int d{\bf r} n({\bf r})\ln \left[\frac{n({\bf r})}{n_0}\right] +
 \frac{U[n]}{k_BT}  .    \label{FEclass}
\end{equation}
The first term on the right hand side of (\ref{FEclass}) is recognizable as the ideal gas free energy $F_{{\rm id}}$ of classical particles, assuming the zero of free energy is set at $\ln \left(n_0 \Lambda^3\right)-1 = 0$, where $\Lambda$ is the de Broglie wavelength. The second term on the right hand side, which is the functional potential, would contain all other effects, including the excluded volume of classical particles. Thus it is identified as $F_{{\rm ex}}$, the free energy in excess of the ideal gas. The functional therefore takes the usual classical DFT form of $F[n] = F_{{\rm id}}[n] + F_{{\rm ex}}[n]$.

\section{Spectral Method}  \label{spectral}

Rather than solving the SCFT equations in real space, one can expand all spatially dependent functions in an infinite basis set $\{f_i({\bf r})\}$ that has the symmetry of the problem encoded in it \cite{Matsen2009, Matsen2006, Matsen1994}. The basis functions should be eigenfunctions of the Laplacian operator and are chosen to be orthonormal according to
\begin{equation}
\frac{1}{V}\int d{\bf r} f_i({\bf r}) f_j({\bf r}) = \delta_{ij}   \label{ortho2}
\end{equation}
where $\delta_{ij}$ is the Kronecker delta. An arbitrary function $g({\bf r})$ is expanded as 
\begin{equation}
g({\bf r}) = \sum_i g_i f_i({\bf r})  .  \label{exp}
\end{equation}
Instead of solving for $g({\bf r})$ at every point in space, one solves instead for a finite number of the coefficients $g_i$, enough to achieve required accuracy. For the set of equations (\ref{w3})-(\ref{init3}), there are also functions of \emph{two} spatial coordinates, in which case a bilinear expansion can be used.
\begin{equation}
g({\bf r},{\bf r}_0) = \sum_{ij} g_{ij} f_i({\bf r}) f_j({\bf r}_0) .  \label{exp2}
\end{equation}

Expanding the single particle partition function (\ref{Q3}), (\ref{exp2}) gives
\begin{eqnarray}
Q &=& \frac{1}{V} \int d{\bf r} q({\bf r},{\bf r},\beta)  \nonumber \\
&=& \frac{1}{V} \sum_{ij} q_{ij}(\beta) \int d{\bf r}  f_i({\bf r}) f_j({\bf r}) \nonumber  \\
&=& \sum_i q_{ii}(\beta)  . \label{Q5}
\end{eqnarray}
The density $n({\bf r})$ (equation (\ref{n3})) can be expanded with either (\ref{exp}) or (\ref{exp2}) to give the two relations
\begin{eqnarray}
n({\bf r}) &=& \sum_i n_i(\beta) f_i({\bf r})   \nonumber \\
&=& \frac{n_0}{Q} \sum_{ij} q_{ij}(\beta) f_i({\bf r}) f_j({\bf r})  .  \label{n8}
\end{eqnarray}
Equating the two expansions, multiplying by $f_k({\bf r})$ and integrating gives
\begin{equation}
n_k(\beta) = \frac{n_0}{Q} \sum_{ij} q_{ij}(\beta) \Gamma_{ijk}  \label{n9}
\end{equation}
where
\begin{equation}
\Gamma_{ijk} = \frac{1}{V} \int d{\bf r} f_i({\bf r})f_j({\bf r})f_k({\bf r})  .  \label{Gamma5}
\end{equation}
The only unspecified quantity in the above equations is $q_{ij}(\beta)$. These components are found by expanding the diffusion equation (\ref{diff3}) using
\begin{equation}
q({\bf r}_0,{\bf r}) = \sum_{ij} q_{ij}(\beta) f_i({\bf r}) f_j({\bf r}_0)   . \label{qij}
\end{equation}
Following the derivation of Matsen \cite{Matsen2009, Matsen2006, Matsen1994}, this gives
\begin{equation}
\frac{d}{d\beta} q_{nm}(\beta) = \sum_j A_{nj} q_{mj}(\beta)   \label{specdiff}
\end{equation}
where 
\begin{equation}
A_{ij} \equiv \frac{\hbar^2}{2m} \varepsilon_i \delta_{ij} - \sum_k w_k \Gamma_{ijk}  \label{Aij}
\end{equation}
where $w_k$ are the expansion coefficients of the field $w({\bf r})$ from (\ref{exp}), and where $\varepsilon_i$ are the eigenvalues of the Laplacian operator with respect to the basis set $\{f_i({\bf r})\}$.
Following Matsen \cite{Matsen2009, Matsen2006, Matsen1994}, equation (\ref{specdiff}) can be solved analytically to give
\begin{equation}
q_{nm}(\beta) = \sum_k e^{A_{nk}\beta}q_{km}(0)   \label{specsolve}
\end{equation}
where $q_{km}(0)$ are the expansion coefficients of the initial condition (\ref{init3}), which gives $q_{km}(0) = \delta_{km}$. Therefore (\ref{specsolve}) is
\begin{equation}
q_{nm}(\beta) = e^{A_{nm}\beta}  .  \label{specsolve2}
\end{equation}
Matsen has discussed how to solve this exponential of a matrix \cite{Matsen2009}. Equation (\ref{specsolve2}) is equivalent to
\begin{equation}
q_{nm}(\beta) = \sum_l U_{nl} e^{\lambda_l \beta} U_{lm}    \label{specsolve3}
\end{equation}
where $\lambda_l$ are the eigenvalues of the matrix $A$ and the columns of $U$ are the normalized eigenvectors of $A$. Thus the problem of solving the set of diffusion equations (\ref{diff3}) is reduced to finding the eigenvalues and eigenvectors of the matrix $A$ of (\ref{Aij}). For a linear polymer, represented by a single diffusion equation, one must solve the same matrix. Thus the computational burden for solving the set of diffusion equations (\ref{diff3}) is, spectrally, the same as solving a single diffusion equation.

\section{Spherical Bessel Expansions and Potential Terms}   \label{sphB}

In order to solve the SCFT equations spectrally, certain quantities need to be expanded in terms of the basis set. Here the basis set is given by (\ref{sphbessel}), which are eigenfunctions of the Laplacian operator, with eigenvalues given by
\begin{equation}
\lambda_n = -\left(\frac{n\pi}{R}\right)^2   .  \label{sphBeig}
\end{equation}
For this basis set, the $\Gamma_{ijk}$ tensor (\ref{Gamma5}) will be
\begin{equation}
\Gamma_{ijk} = -\frac{1}{2} \sqrt{\frac{2}{3}}\left\{{\rm Si}\left[(i+j+k)\pi\right] + {\rm Si}\left[(i-j-k)\pi\right] + {\rm Si}\left[(-i+j-k)\pi\right] + {\rm Si}\left[(-i-j+k)\pi\right]\right\}  \label{Gamma6}
\end{equation}
where ${\rm Si}(x)$ is the Sine Integral given by
\begin{equation}
{\rm Si}(x) = \int_0^{x} \frac{\sin y}{y} dy .   \label{Si}
\end{equation}
Note that (\ref{Gamma6}) is independent of the spherical box size $R$ or other system parameters, and so is universal for all systems with spherical symmetry. Thus it can be determined once and stored if desired. From (\ref{sphBeig}) and (\ref{Gamma6}), the matrix $A_{ij}$ given by (\ref{Aij}) can be specified for a given set of field coefficients $w_i$. 

The field coefficients $w_i$ are determined by the potentials acting on the quantum particles. The hydrogen atom is particularly simple, since there is only one potential, the ``external'' potential, which is the Coulomb interaction between an electron and the nucleus. The free energy contribution of the external field is
\begin{equation}
U_{{\rm ext}}[n] = -N \int \int d{\bf r} d{\bf r}^\prime n({\bf r}) \mathcal{V}(|{\bf r}-{\bf r}^\prime|) \rho_{{\rm ion}}({\bf r}^\prime)  \label{Uext}
\end{equation}
where $\mathcal{V}(r)$ is the Coulomb potential, $\rho_{{\rm ion}}({\bf r})$ is the ionic distribution and $N$ is the atomic number (number of electrons). Atomic units are now being used: $\hbar =1$, $m_e = 1$, $1/4\pi \epsilon_0 = 1$ where $m_e$ is the electron mass and $\epsilon_0$ is the permittivity of free space.  Applying relation (\ref{w3}), the external potential will be
\begin{equation}
w_{{\rm ext}}({\bf r}) = -N \int d{\bf r}^\prime \mathcal{V}(|{\bf r}-{\bf r}^\prime|) \rho_{{\rm ion}}({\bf r}^\prime) . \label{wext}
\end{equation}
In terms of (\ref{wext}), equation (\ref{Uext}) can be written as 
\begin{equation}
U_{{\rm ext}}[n] = N \int d{\bf r} n({\bf r}) w_{\rm ext}({\bf r})  \label{Uext2}
\end{equation}
giving a spectral expansion of
\begin{equation}
U_{{\rm ext}}[n] = V \sum_i n_i w^{{\rm ext}}_i   \label{Uext3}  
\end{equation}
where $n_i$ are the components of the electron density and $w^{{\rm ext}}_i$ are the components of the ion potential. To find the components of $w^{{\rm ext}}_i$, one notes that, ignoring surface terms (due to the finite spherical box of radius $R$), the integral expression (\ref{wext}) can be replaced with the Poisson equation
\begin{equation}
\nabla^2 w({\bf r}) = -4\pi \rho({\bf r})  \label{Poisson}
\end{equation}
as is often done in electrostatics problems. Equation (\ref{Poisson}) is readily expanded in terms of orthonormal basis functions and, using the eigenvalues (\ref{sphBeig}) one finds
\begin{equation}
w_i = 4\pi \rho_i \left(\frac{R}{i\pi}\right)^2  . \label{wgeneral}
\end{equation}
For atomic systems, the ion density distribution will be a Dirac delta function centred at the origin, so
\begin{equation}
\rho_i = \frac{1}{V} \sqrt{\frac{2}{3}}i\pi  \label{rhogeneral}
\end{equation}
where $V = 4\pi R^3/3$ is the size of the finite spherical box. The external potential components are therefore
\begin{equation}
w_i^{{\rm ext}} = -\frac{\sqrt{6}}{Ri\pi}  .  \label{wext2}
\end{equation}
The free energy of the hydrogen atom, expressed in terms of basis functions, will be, from (\ref{FE3}), simply
\begin{equation}
F = -\frac{1}{\beta}\ln Q  - \frac{1}{R}   \label{FEhyd}
\end{equation}
with the external potential $U_{{\rm ext}}[n]$ cancelling with the last term on the right hand side of (\ref{FE3}). The term $1/R$ subtracted on the right hand side of (\ref{FEhyd}) is due to the ignored surface term arising from the solution of the Poisson equation: the integral expression (\ref{wext}) is only a solution to the Poisson equation for infinite boundary conditions. However, since the charge distribution is centred at the origin, one can easily compute the correction due to the finite boundary at $R$, which is the $1/R$ term.

For helium and higher atomic number atoms, there will be electron-electron interactions and exchange-correlation terms in the potential $U[n]$. The electron-electron term will be a Coulomb potential like the external potential:
\begin{equation}
U_{{\rm ee}}[n] = \frac{1}{2} \int \int d{\bf r} d{\bf r}^\prime n({\bf r}) \mathcal{V}(|{\bf r}-{\bf r}^\prime|) n({\bf r}^\prime)  \label{Uee}
\end{equation} 
with the factor of $1/2$ for double counting. From (\ref{w3}), the electron-electron potential is 
\begin{equation}
w_{{\rm ee}}({\bf r}) =  \int d{\bf r}^\prime \mathcal{V}(|{\bf r}-{\bf r}^\prime|) n({\bf r}^\prime)  \label{wee}
\end{equation}
and so (\ref{Uee}) can be written
\begin{equation}
U_{{\rm ee}}[n] = \frac{1}{2} \int d{\bf r} n({\bf r}) w_{\rm ee}({\bf r})  \label{Uee2}
\end{equation}
These forms are the same as for the external potential, so the Bessel expansion of (\ref{Uee2}) and (\ref{wee}) are
\begin{eqnarray}
U_{{\rm ee}}[n] &=& \frac{V}{2} \sum_i n_i w_i^{{\rm ee}}   \label{Uee3} \\
w_i^{{\rm ee}} &=& 4\pi n_i \left(\frac{R}{i\pi}\right)^2   \label{wee2}
\end{eqnarray}
respectively.

The exchange-correlation functional will be taken to be an exchange only local density approximation, using the formula (\ref{ldax}), in order to compare results with Finzel \cite{Finzel2015}. Applying (\ref{w3}) to equation (\ref{ldax}), one gets the exchange potential
\begin{equation}
w_x({\bf r}) = -\left(\frac{3}{\pi}\right)^{\frac{1}{3}} n({\bf r})^{\frac{1}{3}}   \label{wx}
\end{equation}
with which one can write (\ref{ldax}) as
\begin{equation}
U_x[n] = \frac{3}{4}\int d{\bf r} n({\bf r}) w_x({\bf r}) . \label{ldax2}
\end{equation}
In terms of spherical Bessel coefficients, (\ref{ldax2}) will be
\begin{equation}
U_x[n] = \frac{3}{4}V \sum_i n_i w_i^x  .  \label{ldax3}
\end{equation}
Due to the non-linearity of (\ref{wx}), the coefficients $w_i^x$ have to be determined numerically from the real space formula.

The coefficients $w_i^p$ of the Pauli potential $w_p({\bf r})$ can also be determined numerically from the step-function based potentials suggested by Finzel \cite{Finzel2015}. The functional would be
\begin{equation}
U_p[n] = \int d{\bf r} n({\bf r}) w_p({\bf r}) . \label{Pauli}
\end{equation}
In terms of coefficients, this is 
\begin{equation}
U_p[n] = V \sum_i n_i w_i^p  .  \label{Pauli2}
\end{equation}

Overall, the free energy (\ref{FE3}), in terms of basis function coefficients, becomes
\begin{equation}
F = -\frac{N}{\beta} \ln Q - V \sum_i n_i \left(\frac{1}{2} w_i^{\rm ee} + \frac{1}{4}w_i^x\right) - \frac{N}{R} \label{FE4}
\end{equation}
where the fact that the total field coefficients are given by $w_i = w_i^{\rm ext}+w_i^{\rm ee}+w_i^x+w_i^p$ has been used to simplify the expression. The last term on the right-hand side of (\ref{FE4}) is included to account for the finite boundary correction of the Poisson equation, as previously discussed. Note that $w_i^{\rm ext}$ and $w_i^p$ do not appear in (\ref{FE4}) because they are both independent of electron density. This is natural for the ionic, external potential, but Pauli exclusion should not strictly be represented as an external potential. There exist other choices for the SSB Pauli potential that may address this \cite{Finzel2015b, Finzel2016, Finzel2016b, Finzel2017}.

\bibliography{DFTbibliography4}

\end{document}